\documentclass[osajnl,twocolumn,letter]{revtex4}

\usepackage{amsfonts}
\usepackage{amssymb}
\usepackage{graphicx}

\DeclareRobustCommand{\baselinestretch{1}}

\def\ocis#1{
   \noindent \hskip.35in\parbox{5.75in}{\bf  OCIS codes: \rm #1 \hfil } \\ \normalsize \vskip-.1in}

\begin{document}

\title{The complete modulational instability 
gain spectrum of nonlinear QPM gratings}

\author{Joel F. Corney}

\affiliation{Department of Physics, University of Queensland, 
             St. Lucia Qld 4072, Australia}

\author{Ole Bang}

\affiliation{Research Centre COM, Technical University of Denmark, 
         DK-2800 Kongens Lyngby, Denmark}

\begin{abstract}
We consider plane waves propagating in quadratic nonlinear slab waveguides
with nonlinear quasi-phase-matching gratings. 
We predict analytically and verify numerically the complete gain spectrum 
for transverse modulational instability, including hitherto undescribed 
higher order gain bands.
\ocis{190.4410, 190.4420}
\end{abstract}

\maketitle

\section{Introduction}
With the maturing of the quasi-phase-matching (QPM) technique, in particular 
by electric-field poling of ferro-electric materials, such as LiNbO$_3$ 
\cite{Fej98}, by poling of polymers \cite{polymer} and quantum-well 
disordering in semiconductors \cite{Hel00}, the number of applications of 
quadratic nonlinear (or $\chi^{(2)}$) materials has increased significantly. 
Even complicated QPM grating structures are now commercially available
in periodically poled LiNbO$_3$ (PPLN). 
It is therefore more important than ever to have complete knowledge of the 
effects a QPM grating has on the properties of $\chi^{(2)}$ materials.
The most fundamental effect of a QPM grating, with a certain spectrum
of spatial wave vectors, is to allow noncritical phase matching at all 
wavelengths for which the wave-vector mismatch $\Delta k$ matches a 
component of the grating spectrum.
Thus QPM gratings allow for efficient multiple wavelength second
harmonic generation (SHG) \cite{QPMmultiSHG}, which may, for example, be
used for multiple-channel wavelength conversion \cite{ChoParFejBre99}.

In addition to providing phase matching, QPM gratings have the basic effect 
that they induce an asymmetric cubic nonlinearity (ACN) in the equations 
for the average field in the form of self- and cross-phase modulation terms 
\cite{ClaBanKiv97}. 
This ACN appears in linear and/or nonlinear periodic QPM gratings of 
arbitrary shape \cite{CorBan01sol}, it can be focusing or defocusing, 
depending on the sign of the phase mismatch \cite{CorBan01sol}, and its
strength can be increased (e.g., dominating the Kerr nonlinearity) by 
modulation of the grating \cite{BanClaChrTor99}.
In continuous-wave operation the ACN induces an intensity-dependent phase 
mismatch, just as the inherent Kerr nonlinearity, with potential use 
in switching applications \cite{BanGraCor01,KobLedBanKiv98}.  
The ACN further explains a wide spectrum of novel fundamental 
properties of solitons \cite{ClaBanKiv97,CorBan01sol,JohCarTorBan02} and 
modulational instability (MI) \cite{CorBan01mi:prl,CorBan01mi:josab}. 
Importantly the ACN is a general effect of 
non-phase-matched wave interaction and as such appear also in 
homogeneous $\chi^{(2)}$ materials in the cascading limit.  
In fact, in this case the asymmetric signature of the ACN may be measured
as the difference between the properties in upconversion and downconversion, 
since there is no effective $\chi^{(2)}$ nonlinearity competing with the ACN 
as in QPM gratings. 
Such an experiment was recently reported \cite{Trap_ACN} and thus the ACN has
now been confirmed both numerically and experimentally.

In this paper, we present the results of a complete study into the modulational instability of beams in 1+1D  QPM waveguides.  We find that in general, the MI gain spectrum has a multipeaked structure of up to three fundamental bands with accompanying overtones.  Our previous work\cite{CorBan01mi:prl,CorBan01mi:josab} concentrated on the structure of the fundamental  (long-wave instability) bands at low transverse wavenumbers.  These bands are due to MI of the averaged field, and are predicted by averaged equations, provided the effective ACNs are taken into account for the most accurate description.  For example, for QPM with a simultaneous linear and nonlinear grating and/or with a 
nonlinear grating with a dc-value (as with QPM in polymers 
\cite{polymer} and semiconductors \cite{Hel00}), the ACNs can suppress these fundamental bands, making plane waves modulationally stable over hundreds of diffraction 
lengths \cite{CorBan01mi:prl,CorBan01mi:josab}.  Exact Floquet calculations and numerical simulations confirmed the predictions for the low-frequency bands, but also revealed high-frequency bands not predicted by the averaged theory.
These gain bands were surmised to be related to the 
inherent instability in homogeneous (non-phase-matched) $\chi^{(2)}$ media
\cite{CorBan01mi:prl,CorBan01mi:josab}.  
However, an accurate {\em analytic} description of the first overtone bands was
not derived and the higher overtone bands were not discussed 
at all.

Here we concentrate on  PPLN slab waveguides, which are $\chi^{(2)}$ media with a purely nonlinear
QPM grating with no dc-component.  Although more-general kinds of grating can be analysed with our method, we choose to focus on the simpler case for clarity.  
For example, with this simple QPM grating we do not need to take into 
account the ACN for an accurate description of MI 
\cite{CorBan01mi:prl,CorBan01mi:josab}, however our analytic method does allow for such cubic nonlinearities to be included.
We present the first analytical and numerical description of the complete MI spectrum with all overtones.
We show explicitly that the overtone series are caused by MI
in the rapidly varying components of the propagating fields, 
which are in turn induced by the grating.  
We derive approximate analytic expressions for the positions of the
gain bands and compare them with exact Floquet results and 
direct numerical simulation, to find good agreement.

\section{Method}
We consider a linearly polarized electric field propagating in a lossless 
$\chi^{(2)}$ slab waveguide with a QPM grating under conditions for (the
most efficient) first-order QPM and type I SHG. 
The normalized dynamical equations for the slowly varying envelopes of the 
fundamental $E_1(x,z)$ and second harmonic (SH) $E_2(x,z)$ take the form

\begin{eqnarray}
  \label{field_eqns1}
      i\frac{\partial E_1}{\partial z} + \frac{1}{2}\frac{\partial^2 E_1}{\partial x^2} + 
      \chi(z)E_1^*E_2 \exp(i\beta z) &=& 0 , \\ &&\nonumber \\
      \label{field_eqns2}
 i\frac{\partial E_2}{\partial z} + \frac{1}{4}\frac{\partial^2 E_2}{\partial x^2} + 
      \chi(z)E_1^2 \exp(-i\beta z) &=& 0, 
\end{eqnarray}\vspace*{12pt}  
 
\noindent where the asterisk means complex conjugate. The $x$ and $z$ scales have been 
normalised by $x_0$ and $z_0$=$k_1 x_0^2$, respectively, and 
$\beta$=$(k_2-2k_1)z_0$ is the normalized wave-vector mismatch.  
The nonlinearity is periodic in $z$ with expansion 

\begin{eqnarray}
\chi(z) = \sum_n d_n\exp(i n \kappa z),
\end{eqnarray}\vspace*{12pt}   

\noindent where $d_n$=$d_{-n}^*$ ($\chi$ is real)
and the grating wave number $\kappa$ is defined to be positive.
The grating will force the same periodicity in the propagating
fields. We therefore expand these also:

\begin{eqnarray}
E_1(x,z)&=&\sum_n w_n(x,z) \exp{(in\kappa z)},\\ &&\nonumber \\ 
E_2(x,z)&=&\sum_n v_n(x,z) \exp{(in\kappa z-i\tilde\beta z)},
\end{eqnarray}\vspace*{12pt} 
     
\noindent where $\tilde\beta$=$\beta-s\kappa$ is the residual mismatch, with
$s$=${\rm sign}(\beta)$ for first-order QPM.  
Substituting all expansions into 
Eqs.~(\ref{field_eqns1}-\ref{field_eqns2}) gives

\begin{eqnarray}
  \label{coupled1}
  & & 
      (\hat L_1 -n\kappa) w_n +\sum_{l,m} d_{n+m-l-s}w_m^*v_l = 0, \\ &&\nonumber\\ 
  \label{coupled2}
  & &
      (\hat L_2 -n\kappa) v_n +\sum_{l,m} d_{n-m-l+s}w_mw_l = 0,
\end{eqnarray}\vspace*{12pt}  
    
\noindent where $\hat L_j=i\partial_z+\partial_x^2/(2j)+(j-1)\tilde\beta$.  
This set of equations has plane-wave solutions of the form $w_n(x,z)=\bar w_n 
\exp{(i\Lambda z)}$, $v_n(x,z)=\bar v_n \exp{(2i\Lambda z)}$ 
\cite{CorBan01mi:prl,CorBan01mi:josab}.

To study MI we consider small perturbations of the plane-wave solutions: 

\begin{eqnarray}
w_n(x,z)&=&\left[\bar w_n + \epsilon_n(x,z)\right]\exp{(i\Lambda z)}, \\ &&\nonumber\\ 
v_n(x,z)&=&\left[\bar v_n + \mu_n(x,z)\right]\exp{(i2\Lambda z)}.  
\end{eqnarray}\vspace*{12pt}
      
\noindent Substitution into Eqs.~(\ref{coupled1}-\ref{coupled2}) gives the linearized
equations

\begin{eqnarray}
  \label{coupled_linear1}
       (\hat L_1' -n\kappa) \epsilon_n +\sum_{l,m} d_{n+m-l-s}(\bar w_m^*
       \mu_l + \bar v_l\epsilon_m^*)&  =& 0, \\ &&\nonumber \\ 
  \label{coupled_linear2}   
       (\hat L_2' -n\kappa) \mu_n + 2\sum_{l,m} d_{n-m-l+s}\bar w_m 
       \epsilon_l& =& 0,
\end{eqnarray}\vspace*{12pt}      

\noindent where $\hat{L}_j' = \hat{L}_j - j\Lambda$.  
Writing the perturbations in the form 

\begin{eqnarray}
\epsilon_n(x,z)& = &\delta_n^{(1)}(z)
\exp{(i \nu x)} + \delta_n^{(2)*}(z)\exp{(-i\nu x)},\\ &&\nonumber\\
\mu_n(x,z) & = & \delta_n^{(3)}(z)\exp{(i \nu x)} + \delta_n^{(4)*}(z)\exp{(-i\nu x)},
\end{eqnarray}\vspace*{12pt}
      
\noindent one obtains a linear matrix equation for the 
perturbation $\vec{\delta}_n$=$(\delta_n^{(1)},\delta_n^{(2)},\delta_n^{(3)},
\delta_n^{(4)})^T$, which couples all Fourier components.

\begin{figure}
  \centerline{\scalebox{.29}{\includegraphics{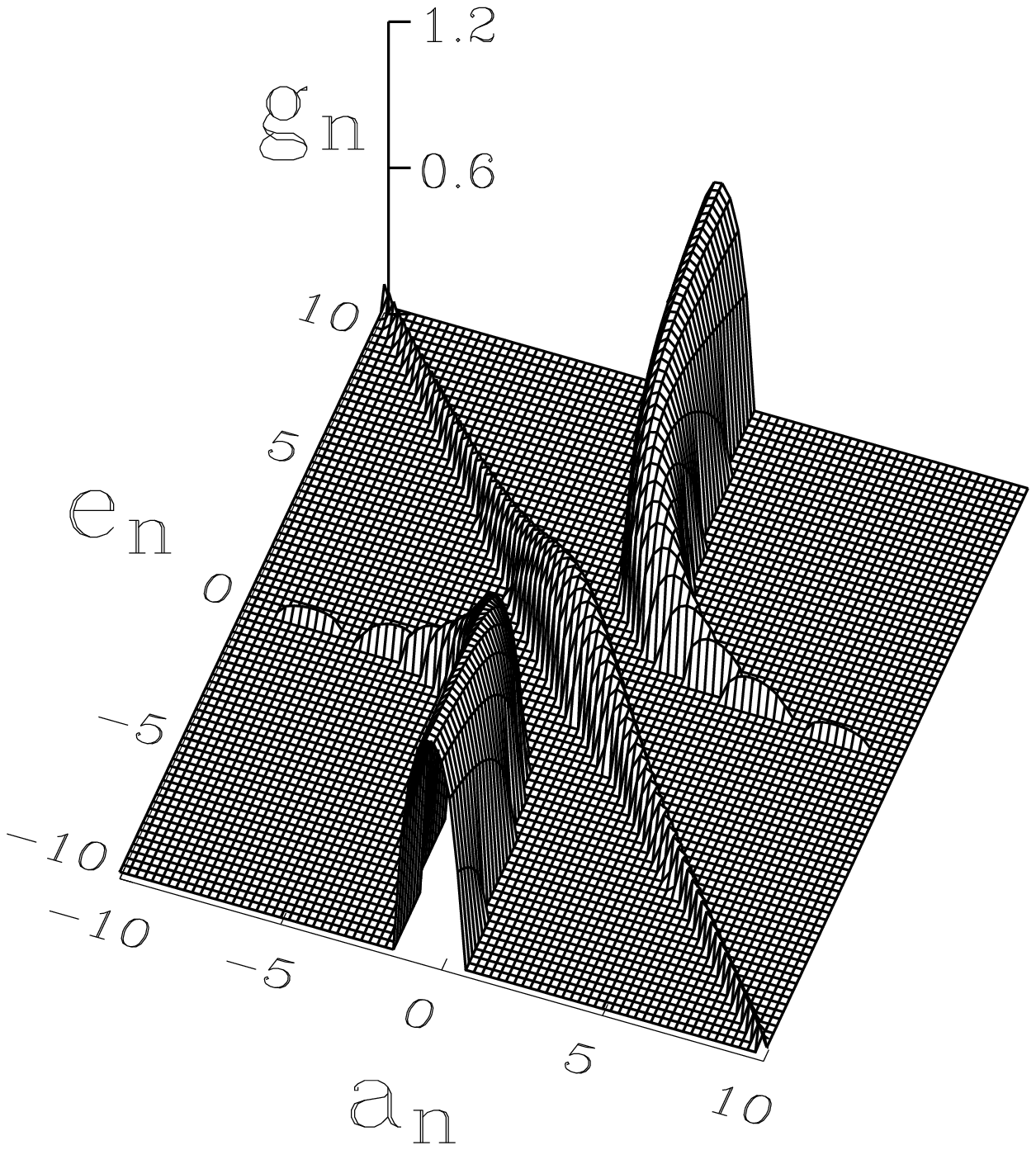}}
              \scalebox{.25}{\includegraphics{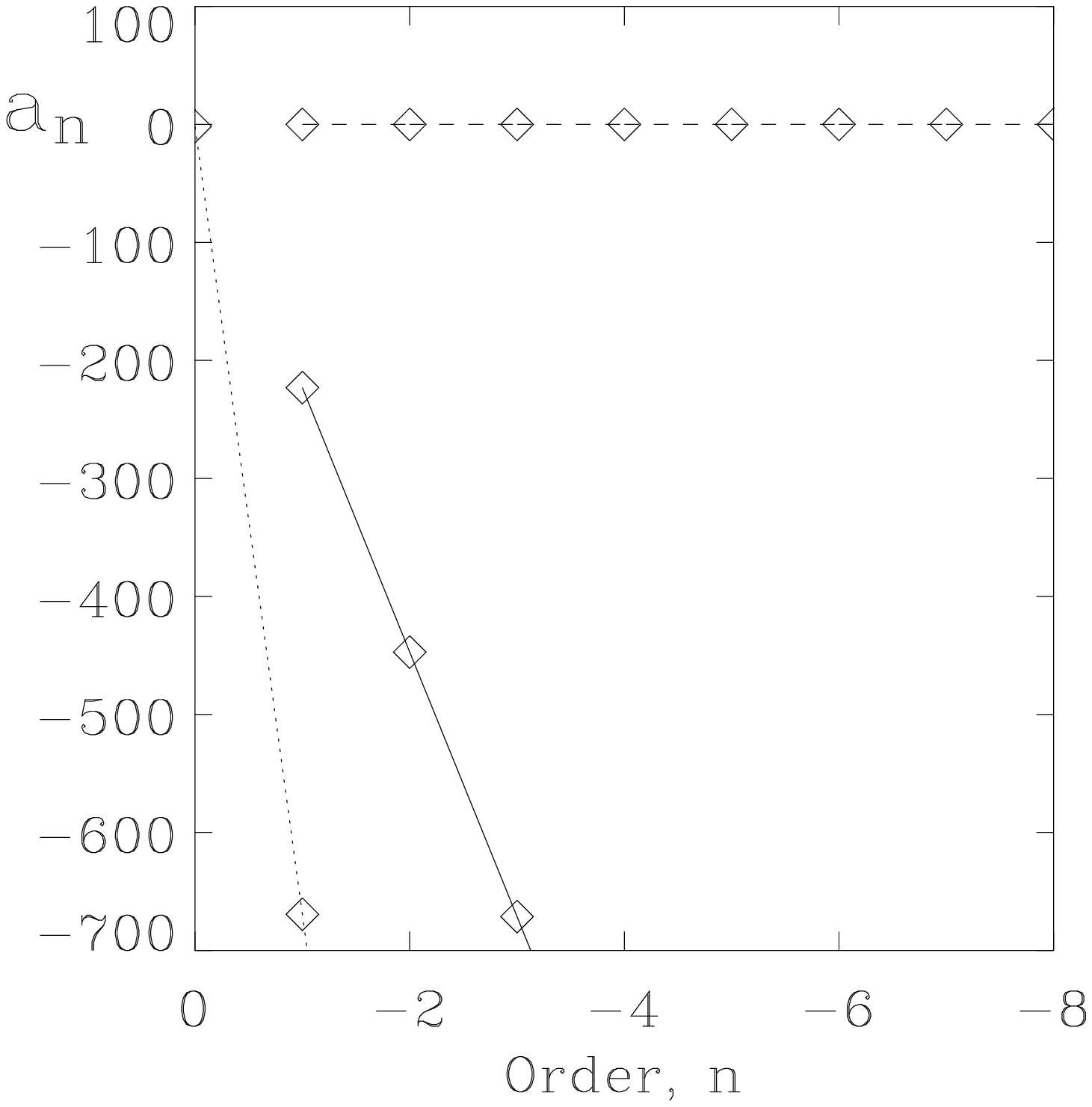}}}
  \caption{Left: Maximum gain versus $a_n$ and $e_n$ for $|b_n|^2$=1 
  and $|c|^2$=2. Right: $a_n(\nu_n)$ for $\nu_n$ real and found from 
  Eq.~(\ref{nudia}) (solid line), from Eq.~(\ref{nuhyp}) with plus 
  sign (dotted line), and Eq.~(\ref{nuhyp}) with minus sign (dashed 
  line), for $\tilde{\beta}$=0, $\Lambda$=1, and $\kappa$=670.}
  \label{branches}
\end{figure}

To derive a simple result we consider nearly phase-matched interaction with
$|\kappa|$$\sim$$|\beta|$$\gg$1 in the typical (for PPLN) square QPM grating, 
for which $d_n$=0 for $n$ even and $d_n$=$2/(i\pi n)$ for $n$ odd. 
The requirement $|\kappa|\sim|\beta|\gg1$ allows us to assume that the 
amplitudes $w_n$ and $v_n$ vary slowly compared to $\exp{(i\kappa z)}$ 
and that the higher harmonics are much smaller than the dc-fields,
$|\bar w_0|$, $|\bar v_0|\gg|\bar w_{n \ne 0}|$, $|\bar v_{n \ne 0}|$.    
In this case the structure of the matrix equation means that the evolution 
of the perturbation in each component is to a good approximation decoupled 
from the other components,

\begin{eqnarray}
  \frac{\partial\vec{\delta}_n}{\partial z} & \approx & i \left[
  \begin{array}{cccc}
    a_n    & b_n  & c   & 0 \\
    -b_n^* & -a_n & 0   & -c^* \\
    2c^*   & 0    & e_n & 0 \\  
    0      & -2c  & 0   & -e_n 
  \end{array} 
  \right]\vec{\delta}_n \equiv M_n \vec{\delta}_n,
  \label{MI_matrix}
\end{eqnarray}\vspace*{12pt}  
    
\noindent where $a_n$=$-\nu^2/2-\Lambda-n\kappa$, $b_n$=$d_{2n-s}\bar{v}_0$,
$c$=$d_{-s}\bar{w}_0^*$, and $e_n$=$-\nu^2/4+\tilde\beta-2\Lambda-n\kappa$.
The eigenvalues of $M_n$ are

\begin{eqnarray}
  \label{evalues}
   \lambda_n^2 & =&  A_n\pm\sqrt{A_n^2-B_n},
\end{eqnarray}\vspace*{12pt} 
     
\noindent where 

\begin{eqnarray}
  A_n & = & \frac{1}{2}(|b_n|^2 - a_n^2 - 4|c|^2 - e_n^2),\\ &&\nonumber \\ 
  B_n & = &  (a_n e_n - 2|c|^2)^2 - e_n^2|b_n|^2. 
\end{eqnarray}\vspace*{12pt}      

\noindent Any positive real part of an eigenvalue corresponds to MI with the gain 
$g_n(\nu) = \Re(\lambda_n)$.

\section{Results}

\begin{figure}
  \centerline{\scalebox{.45}{\includegraphics{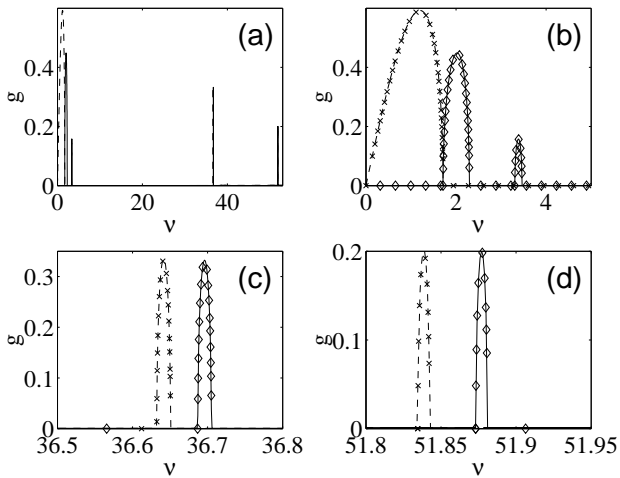}}}
  \caption{MI gain for $\Lambda$=1 (dashed) and $\Lambda$=$-1$ (solid) 
  calculated by Floquet theory. Crosses and diamonds show the theoretical
  prediction (\ref{evalues}). $\kappa$=$\beta$=670.}
  \label{gtheory}
\end{figure}

For the gratings that we consider here, $|c|^2\simeq\Lambda(2\Lambda-\tilde\beta)$ and $|b_n|^2\simeq\Lambda^2/
(2n-s)^2$.  Analysing the gain versus $a_n$ and $e_n$ reveals three gain bands, with 
extrema remaining close to $a_n = -e_n$ (diagonal branch DB) and 
$a_ne_n=2|c|^2$ (hyperbolic branches HB$_+$, HB$_-$), as illustrated in Fig.~\ref{branches}
for $|b_n|^2=1$ and $|c|^2=2$.
The extrema $a_n = -e_n$ of DB correspond to

\begin{eqnarray}
  \nu_n^2 = -8n\kappa/3 - 4\Lambda + 4\tilde\beta/3,
  \label{nudia}
\end{eqnarray}\vspace*{12pt}

\noindent while the extrema $a_ne_n=2|c|^2$ of HB$_+$ and HB$_-$ correspond to

\begin{eqnarray}
 \nu_n^2 = 2\tilde\beta - 3n\kappa - 5\Lambda \pm 
         \sqrt{(2\tilde\beta - n\kappa - 3\Lambda)^2 +16|c|^2}.  
 \label{nuhyp}
\end{eqnarray}\vspace*{12pt}

For $n\ne0$ the HB$_+$ bands appear at $\nu^2 \simeq  -4\kappa n - 8\Lambda + 4\tilde\beta$ and the HB$_-$ bands at 
$\nu^2\simeq-2\kappa n-2\Lambda$.  
Thus we have the structure of up to 3 gain bands in the average field ($n$=0), 
each with a set of equally spaced (in $\nu^2$) weak overtone gain bands ($n\ne 0$). 
However, the gain is large only near $a_n$=0 (see Fig.~\ref{branches}).  For $n\neq 0$, $a_n\simeq0$ can be satisfied without violating the assumptions only on HB$_-$, and thus the overtones of the other branches will be small.

Consider a system at exact phase-matching $\tilde\beta=0$, with dimensionless parameters $|\Lambda|=1$ and $\kappa = 670$.  For a LiNbO$_3$ QPM grating with period $\lambda_g = 17.6 \mu m$ designed to phase-match at wavelengths of around $\lambda =1550 $nm, these dimensionless parameters correspond to having an input intensity of $I=6.7 $GW$/$m$^2$, if one assumes a waveguide depth of $y_0=3.1\mu $m.  The distance scaling parameters for this case are $x_0 = 21.6 \mu $m and $z_0 = 1.89 $mm.  For fixed distance scaling, different values of the magnitude of $\Lambda$ correspond to different input intensities, and different signs of $\Lambda$ correspond to different ratios of the fundamental to second harmonic\cite{CorBan01mi:josab}.  On the other hand, keeping $|\Lambda|$ fixed while varying the input intensity corresponds to changing the distance scaling: to a good approximation, $z_0 \propto I^{-\frac{1}{2}}$, $x_0 \propto I^{-\frac{1}{4}}$, and hence $\kappa\propto I^{\frac{1}{2}}$.

For $\Lambda$=1 our analytic results predict only one fundamental $n$=0 gain 
band (HB$_+$) and three gain bands in all overtones.
However the overtones of DB and HB$_+$ are too small to be seen.  Thus the visible gain bands have maxima at $\nu_0$=1.18
(HB$_+$), $\nu_{-1}$=36.64, $\nu_{-2}$=51.84, etc. (all 
HB$_-$).
In physical units these are  $\nu_0 =0.055 \mu $m$^{-1}$, $\nu_{-1}=1.697 \mu $m$^{-1}$, and $\nu_{-2}=2.401\mu $m$^{-1}$, respectively. The analytical results agree with exact Floquet calculations
\cite{CorBan01mi:prl,CorBan01mi:josab}, as seen in Fig.~\ref{gtheory}.  

\begin{figure}
  \centerline{\scalebox{.45}{\includegraphics{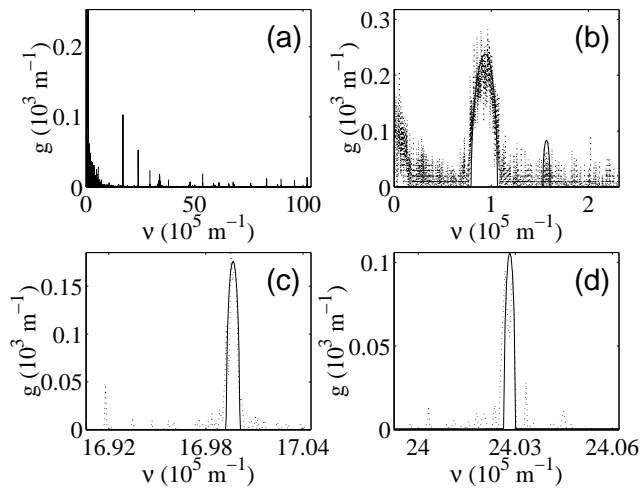}}}
  \caption{Gain calculated by numerical simulation [(a) and dotted line 
  in (b-d)] and by Floquet theory [solid line in (b-d)]. 
  $\kappa=\beta=670$ and $\Lambda=-1$.}
  \label{gnumerical}
\end{figure}

For $\Lambda$=$-1$ two fundamental $n$=0 bands are predicted 
(DB and HB$_+$), with one set of visible overtones from HB$_-$. The maxima are at $\nu_0$=2 (DB), $\nu_0$=3.38 (HB$_+$), $\nu_{-1}$=36.70, 
$\nu_{-2}$=51.88, etc. (all HB$_-$).  In physical units these are  $\nu_0=0.093\mu $m$^{-1}$, $\nu_0=0.0157\mu $m$^{-1}$, $\nu_{-1}=1.700\mu $m$^{-1}$, and $\nu_{-2}=2.403\mu $m$^{-1}$, respectively.
This is again confirmed by the exact Floquet calculations seen in
Fig.~\ref{gtheory}.
Figure \ref{gtheory} also shows the gain profiles predicted by 
Eq.~(\ref{evalues}), which agree well with the exact Floquet results.
Direct numerical simulation \cite{CorBan01mi:prl,CorBan01mi:josab} of 
Eqs.~(\ref{field_eqns1}-\ref{field_eqns2}) with MI seeded by noise also
confirmed our results (see Fig.~\ref{gnumerical}).

We see from Eqs.\ (\ref{nudia}-\ref{nuhyp}) that the dimensionless spectral positions of the higher-order bands depends on $\kappa$, whereas the fundamental bands do not.  The different dependence implies a different scaling with input intensity, since for fixed $|\Lambda|$, both the distance scalings $x_0$, $z_0$ and the dimensionless wave vector $\kappa$  depend on the intensity.  The result is that the spatial frequencies of the higher-harmonic instabilities remain the same, in physical units, whereas the frequencies of the fundamental-band instabilities generally increase with the quartic root of the intensity.

\section{Summary}
In conclusion, we have presented a simple theory able to accurately predict
the complete MI spectrum in general QPM gratings in $\chi^{(2)}$ 
materials. In particular, we have predicted and verified that overtone 
gain bands originate from MI in the higher-order Fourier components of 
the field.
This research is supported by the Danish Technical Research Council 
(Grant No. 26-00-0355) and the Australian Research Council.


\begin{thebibliography}{}
\bibitem{Fej98}
   M.M. Fejer, in \emph{Beam Shaping and Control with Nonlinear Optics},
   eds. F. Kajzar and R. Reinisch, 375--406 (Plenum, New York, 1998).

\bibitem{polymer}
   V. Ricci, G.I. Stegeman, and K.P. Chan,
   ``Poling of multilayer polymer films for modal dispersion phase
   matching of second-harmonic generation: effects in glass-transition
   temperature matching in different layers",
   J. Opt. Soc. Am. B. {\bf 17}, 1349--1353 (2000).

\bibitem{Hel00}
   A. Saber Helmy, D.C. Hutchings, T.C. Kleckner, J.H. Marsh, A.C. Bryce,
   J.M. Arnold, C.R. Stanley, J.S. Aitchison, C.T.A. Brown, K. Moutzouris,
   and M. Ebrahimzadeh,
  ``Quasi-phase-matching in GAAS-ALAS superlattice waveguides via bandgap
   tuning using quantum well intermixing",
   \ol {\bf 25}, 1370--1372 (2000).

\bibitem{QPMmultiSHG}
   P. Baldi, C.G. Trevino-Palacios, G.I. Stegeman, M.P. De Micheli, 
   D.B. Ostrowsky, D. Delacourt, and M. Papuchon,
   ``Simultaneous generation of red, green and blue light in room
   temperature periodically poled lithium niobate waveguides using
   single source",
   Electron. Lett. {\bf 31}, 1350--1351 (1995).

\bibitem{ChoParFejBre99}
   M.H. Chou, K.R. Parameswaran, M.M. Fejer, and I. Brener,
   ``Multiple-channel wavelength conversion by use of engineered
   quasi-phase-matching structures in LiNbO$_3$ waveguides",
   \ol {\bf 24}, 1157--1159 (1999).

\bibitem{ClaBanKiv97}
   C. Balslev Clausen, O. Bang, and Y.S. Kivshar, 
   ``Spatial solitons and induced Kerr effects in quasi-phase-matched 
   quadratic media",
   \prl {\bf 78}, 4749--4752 (1997).

\bibitem{CorBan01sol}
   J.F. Corney and O. Bang,
   ``Solitons in quadratic nonlinear photonic crystals",
   \pre {\bf 64}, 047601-1--047601-4 (2001).

\bibitem{BanClaChrTor99}
   O. Bang, C. Balslev Clausen, P.L. Christiansen, and L. Torner,
   ``Engineering competing nonlinearities",
   \ol {\bf 24}, 1413--1415 (1999).

\bibitem{BanGraCor01}
   O. Bang, T.W. Graversen, and J.F. Corney, 
   ``Accurate switching intensities and optimal length scales in 
   quasi-phase-matched materials", 
   \ol {\bf 26}, 1007--1009 (2001).

\bibitem{KobLedBanKiv98}
   A. Kobyakov, F. Lederer, O. Bang, and Y.S. Kivshar, 
   ``Nonlinear phase shift and all-optical switching in 
   quasi-phase-matched quadratic media", 
   \ol {\bf 23} 506--508(1998).

\bibitem{JohCarTorBan02}
   S.K. Johansen, S. Carrasco, L. Torner, and O. Bang,
   ``Engineering of spatial solitons in two-period QPM structures",
   \oc {\bf 203}, 393--402 (2002).

\bibitem{CorBan01mi:prl}
   J.F. Corney and O. Bang, 
   ``Modulational instability in periodic quadratic nonlinear materials", 
   \prl {\bf 87}, 133901-1--133901-4 (2001);

\bibitem{CorBan01mi:josab}
   J.F. Corney and O. Bang, 
   ``Plane waves in periodic, quadratically nonlinear slab waveguides: 
   stability and exact Fourier structure",
   \josab {\bf 19}, 812--821 (2002).

\bibitem{Trap_ACN}
   P. Di Trapani, A. Bramati, S. Minardi, W. Chinaglia, C. Conti, 
   S. Trillo, J. Kilius, and G. Valiulis, 
   ``Focusing versus defocusing nonlinearities due to parametric 
   wave mixing"
   \prl {\bf 87}, 183902-1--183902-4 (2001).
\end{thebibliography}
\end{document}